\documentstyle[12pt,fleqn]{article}

\font\twlgot =eufm10 scaled \magstep1
\font\egtgot =eufm8
\font\sevgot =eufm7

\font\twlmsb =msbm10 scaled \magstep1
\font\egtmsb =msbm8
\font\sevmsb =msbm7

\newfam\gotfam
\def\pgot{\fam\gotfam\twlgot}
\textfont\gotfam\twlgot
\scriptfont\gotfam\egtgot
\scriptscriptfont\gotfam\sevgot
\def\got{\protect\pgot}

\newfam\msbfam
\textfont\msbfam\twlmsb
\scriptfont\msbfam\egtmsb
\scriptscriptfont\msbfam\sevmsb
\def\Bbb{\protect\pBbb}
\def\pBbb{\relax\ifmmode\expandafter\Bb\else\typeout{You cann't use
Bbb in text mode}\fi}
\def\Bb #1{{\fam\msbfam\relax#1}}

\newcommand{\gO}{{\got O}}
\newcommand{\gQ}{{\got Q}}

\def\thebibliography#1{\section*{
References}\list
  {\arabic{enumi}.}{\settowidth\labelwidth{#1}\leftmargin\labelwidth
    \advance\leftmargin\labelsep
    \usecounter{enumi}}
    \def\newblock{\hskip .11em plus .33em minus .07em}
    \sloppy\clubpenalty4000\widowpenalty4000
    \sfcode`\.=1000\relax}

\let\Large=\large

\def\op#1{\mathop{\fam0 #1}\limits}

\newcommand{\beq}{\begin{equation}}
\newcommand{\eeq}{\end{equation}}
\newcommand{\ben}{\begin{eqnarray}}
\newcommand{\een}{\end{eqnarray}}
\newcommand{\be}{\begin{eqnarray*}}
\newcommand{\ee}{\end{eqnarray*}}
\newcommand{\bea}{\begin{eqalph}}
\newcommand{\eea}{\end{eqalph}}

\newcommand{\cO}{{\cal O}}
\newcommand{\cQ}{{\cal Q}}

\newcommand{\la}{\lambda}
\newcommand{\La}{\Lambda}
\newcommand{\f}{\phi}

\newcommand{\m}{\mu}

\newcommand{\G}{\Gamma}

\newcommand{\th}{\theta}

\newcommand{\si}{\sigma}
\newcommand{\w}{\wedge}

\newcommand{\wh}{\widehat}
\newcommand{\ol}{\overline}
\newcommand{\dr}{\partial}

\newcommand{\ar}{\op\longrightarrow}

\newcounter{eqalph}
\newcounter{equationa}
\newcounter{example}
\newcounter{remark}
\newcounter{theorem}
\newcounter{proposition}
\newcounter{lemma}
\newcounter{corollary}
\newcounter{definition}

\def\theremark{\arabic{remark}}

\def\thedefinition{\arabic{definition}}

\newenvironment{proof}{\noindent {\it Proof.}}{\hfill $\Box$
\medskip }

\newenvironment{prop}{\refstepcounter{definition} \medskip\noindent
PROPOSITION \thedefinition.\it}{\medskip }
\newenvironment{lem}{\refstepcounter{definition} \medskip\noindent  LEMMA
\thedefinition.\it }{\medskip }

\newenvironment{eqalph}{\stepcounter{equation}
\setcounter{equationa}{\value{equation}}
\setcounter{equation}{0}

\begin{eqnarray}}{\end{eqnarray}\setcounter{equation}{\value{equationa}}}

\begin{document}
\hbox{}

{\parindent=0pt 

{ \Large \bf On the Global Calculus in Local Cohomology in BRST Theory}
\bigskip

{\sc G.Giachetta$^\dagger$\footnote{giachetta@campus.unicam.it},
L.Mangiarotti$^\dagger$\footnote{mangiaro@camserv.unicam.it}  and
G.Sardanashvily$^\ddagger$}\footnote{sard@campus.unicam.it;
sard@grav.phys.msu.su} 

{ \small

{\it $^\dagger$ Department of Mathematics and Physics, University of Camerino,
62032 Camerino (MC), Italy

$^\ddagger$ Department of Theoretical Physics,
Physics Faculty, Moscow State University, 117234 Moscow, Russia}
\bigskip

{\bf Abstract}. We show that the cohomology groups of the horizontal (total)
differential on horizontal (local) exterior forms on the infinite-order jet
manifold of an affine bundle coincide with the De Rham cohomology groups of the
base manifold. This prevents one from the topological obstruction to
definition of global descent equations in BRST theory on an arbitrary affine
bundle. }

\section{Introduction}

Let $Y\to X$ be a smooth fibre bundle of some classical field model.
We study cohomology of exterior forms on the infinite-order jet space
$J^\infty Y$ of $Y\to X$. This cohomology plays an important role in
the field-antifield BRST formalism for constructing the descent equations
[1-4]. 

In the framework of this BRST formalism, 
one considers the so-called horizontal complex
\beq
0\to\Bbb R\to \cO^0_\infty \ar^{d_H}\cO^{0,1}_\infty\ar^{d_H}\cdots  
\op\longrightarrow^{d_H} 
\cO^{0,n}_\infty,  \label{+481}
\eeq
where $\cO^{0,*}_\infty$ is a subalgebra of horizontal (semibasic)
exterior forms on $J^\infty Y$, and $d_H$ is the horizontal (total)
differential. Extended to the jet
space of ghosts and antifields, these forms  and their cohomology are called
local forms and local cohomology. Given the BRST operator
${\bf s}$, one defines the total BRST operator ${\bf s} +d_H$ and examines the
BRST cohomology modulo $d_H$. 

It should be emphasized that, the above mentioned BRST formalism is
formulated on a contractible fibre bundle $Y=\Bbb R^{n+m}\to \Bbb R^n$.
Of course, this is not the generic case of gauge theory and its outcomes
to topological field models and anomalies. The key point is that, in
this case, the horizontal complex (\ref{+481}) is exact. This fact called the
algebraic Poincar\'e lemma is crucial for constructing the (local) descent
equations in BRST theory. To write global descent equations in a non-trivial
topological context, one should study (global) cohomology of the complex
(\ref{+481}).  The question is that there are (at least) two classes of
exterior forms on the infinite-order jet space
$J^\infty Y$, and both of them fail to be differential forms on $J^\infty Y$
in a rigorous sense because $J^\infty Y$ is not a Banach manifold [5-7]. 

Recall that the infinite-order jet space of a smooth fibre bundle $Y\to X$ is
defined as a projective limit $(J^\infty Y,\pi^\infty_j)$ of the surjective
inverse system
\beq
X\op\longleftarrow^\pi Y\op\longleftarrow^{\pi^1_0}\cdots \longleftarrow
J^{r-1}Y \op\longleftarrow^{\pi^r_{r-1}} J^rY\longleftarrow\cdots 
\label{5.10}
\eeq
of finite-order jet manifolds $J^rY$ [5-8]. Provided with
the projective limit topology, $J^\infty Y$ is a paracompact Fr\'echet (but
not Banach)  manifold \cite{tak2,bau,abb}. Its paracompactness will be
an essential tool for cohomological calculations below.
Given a bundle coordinate chart
$(\pi^{-1}(U_X); x^\la,y^i)$ on the fibre bundle $Y\to X$, we have the
coordinate chart
$((\pi^\infty)^{-1}(U_X); x^\la, y^i_\La)$, $0\leq|\La|,$ on $J^\infty Y$,
together with the transition functions  
\beq
{y'}^i_{\la+\La}=\frac{\dr x^\m}{\dr x'^\la}d_\m y'^i_\La, \label{55.21}
\eeq
where $\La=(\la_k...\la_1)$, $|\La|=k$,
is a multi-index, 
 $\la+\La$ is the multi-index $(\la\la_k\ldots\la_1)$ and
$d_\la$ are the total derivatives 
\be
d_\la = \dr_\la + \op\sum_{|\La|=0} y^i_{\la+\La}\dr_i^\La.
\ee

The differential calculus on $J^\infty Y$ can be
introduced as operations on the $\Bbb R$-ring $\cQ^0_\infty$ of
locally pull-back functions on
$J^\infty Y$. A real function
$f$ on $J^\infty Y$ is called so if, for each
point $q\in J^\infty Y$, there is a neighbourhood $U_q$ such
that $f|_{U_q}$ is the pull-back of a smooth function on some 
finite-order jet manifold
$J^kY$ with respect to the surjection $\pi^\infty_k$. It should be emphasized
that the paracompact space 
$J^\infty Y$ admits the partition of unity performed by elements of
$\cQ^0_\infty$ \cite{tak2,bau}. The difficulty lies in the geometric
interpretation of derivations of the $\Bbb R$-ring $\cQ^0_\infty$ as vector
fields on the Fr\'echet manifold $J^\infty Y$ and their dual as differential
forms on
$J^\infty Y$ \cite{tak1}. 

Therefore, one usually considers the 
subring
$\cO^0_\infty$ of the ring $\cQ^0_\infty$ which consists of the
pull-back onto $J^\infty Y$ of smooth functions on finite-order jet spaces.
The Lie algebra of derivations of $\cO^0_\infty$ is isomorphic to the
projective limit onto $J^\infty Y$ of the Lie algebras of projectable vector
fields on finite-order jet manifolds. The associated algebra of differential
forms is introduced as the direct limit $(\cO^*_\infty, \pi^{\infty*}_k)$ of
the direct system  
\beq
\cO^*(X)\op\longrightarrow^{\pi^*} \cO^*(Y) 
\op\longrightarrow^{\pi^1_0{}^*} \cO_1^*
\op\longrightarrow^{\pi^2_1{}^*} \cdots \op\longrightarrow^{\pi^r_{r-1}{}^*}
 \cO_r^* \longrightarrow\cdots \label{5.7}
\eeq
of differential $\Bbb R$-algebras  $\cO^*_r$ of exterior forms on finite-order
jet manifolds $J^rY$. This direct limit exists in the category of
$\Bbb R$-modules, and the direct limits of the familiar
operations on exterior forms make $\cO^*_\infty$ a differential
exterior $\Bbb R$-algebra. This algebra consists of
all exterior forms on finite-order jet manifolds modulo the pull-back
identification. Therefore, one usually thinks of elements of $\cO^*_\infty$ as
being the pull-back onto $J^\infty Y$ of exterior forms on finite-order jet
manifolds. 
Being restricted to a coordinate chart $(\pi^\infty)^{-1}(U_X)$ on $J^\infty
Y$, elements of 
$\cO^*_\infty$ 
can be written in the familiar coordinate form, where basic forms
$\{dx^\la\}$ and contact 1-forms
$\{\th^i_\La=dy^i_\La -y^i_{\la+\La}dx^\la\}$ provide 
the local generators of the algebra $\cO_\infty^*$. There is
the canonical splitting of the space of $m$-forms
\be
\cO^m_\infty =\cO^{0,m}_\infty\oplus
\cO^{1,m-1}_\infty\oplus\ldots\oplus\cO^{m,0}_\infty
\ee
into spaces $\cO^{k,m-k}_\infty$ of $k$-contact forms. 
Accordingly, the
exterior differential on $\cO_\infty^*$ is
decomposed into the sum $d=d_H+d_V$
of horizontal and vertical differentials
\be
&& d_H:\cO_\infty^{k,s}\to \cO_\infty^{k,s+1}, \qquad d_H(\f)= dx^\la\w
d_\la(\f), \qquad \f\in\cO^*_\infty,\\ 
&& d_V:\cO_\infty^{k,s}\to \cO_\infty^{k+1,s}, \qquad
d_V(\f)=\th^i_\La \w \dr_\La^i\f,
\ee
which obey the nilpotency rule
\beq
d_H\circ d_H=0, \qquad d_V\circ d_V=0, \qquad d_V\circ d_H
+d_H\circ d_V=0. \label{lmp50}
\eeq

In studying the algebra $\cO^*_\infty$ of pull-back exterior
forms on $J^\infty Y$, the key point of
 lies in the fact that
the infinite-order De Rham complex of these forms 
 \beq
0\to \Bbb R\to
\cO^0_\infty\op\longrightarrow^d\cO^1_\infty\op\longrightarrow^d
\cdots
\label{5.13}
\eeq
is the direct limit of the De Rham complexes of exterior forms on
finite-order jet manifolds. Then, as was repeatedly proved,
the cohomology groups $H^*(\cO^*_\infty)$ of the complex (\ref{5.13}) are equal
to the De Rham cohomology groups
$H^*(Y)$ of the fibre bundle $Y$ \cite{tak2,bau}. This fact enables one
to say something on the topological obstruction to the exactness of the
(infinite-order) variational complex in the calculus of variations
in field theory [6,7,10-12]. At the same time, the
$d_H$-cohomology of the horizontal complex (\ref{+481}) of
pull-back exterior forms on $J^\infty Y$ remains unknown. The local exactness
of this complex only has been repeatedly proved (see, e.g.,
\cite{tul,olver}). If a fibre bundle $Y\to X$ admits a global section, there
is also a monomorphism of the De Rham cohomology groups
$H^*(X)$ of the base $X$ to the cohomology groups of the complex
(\ref{+481}) \cite{jpa00}.

Here, we show that the problem of cohomology of the horizontal complex 
has a comprehensive solution by enlarging the algebra 
$\cO^*_\infty$ to the algebra $\cQ^*_\infty$ of the above mentioned locally
pull-back exterior forms on $J^\infty Y$. 
We introduce these forms in an algebraic way as global sections of the sheaf
$\gQ^*_\infty$ of differential algebras on
$J^\infty Y$ which is the direct limit of the direct system 
\beq
\gO^*_X\op\longrightarrow^{\pi^*} \gO^*_Y 
\op\longrightarrow^{\pi^1_0{}^*} \gO_1^*
\op\longrightarrow^{\pi^2_1{}^*} \cdots \op\longrightarrow^{\pi^r_{r-1}{}^*}
 \gO_r^* \longrightarrow\cdots \label{5.7'}
\eeq
of sheaves of exterior
forms on finite-order jet manifolds $J^rY$.
As a consequence, we
have the exact sequence of sheaves
\beq
0\to \Bbb R \to \gQ^0_\infty \ar^{d_H} \gQ^{0,1}_\infty
\ar^{d_H}\cdots  
\op\longrightarrow^{d_H} 
\gQ^{0,n}_\infty \label{lmp92}
\eeq
of horizontal forms on $J^\infty Y$ and the corresponding complex of
$\cQ^0_\infty$-modules of their global sections
\beq
0\to \Bbb R \to \cQ^0_\infty \ar^{d_H} \cQ^{0,1}_\infty
\ar^{d_H}\cdots  
\op\longrightarrow^{d_H} 
\cQ^{0,n}_\infty. \label{lmp94}
\eeq
Since $J^\infty Y$ is paracompact and admits a partition of
unity by elements of $\cQ^0_\infty$, all sheaves $\gQ^{0,m}_\infty$
in the exact sequence (\ref{lmp92}) are fine and, consequently, acyclic.
Therefore, the well-known theorem on a resolution of a sheaf \cite{hir} can be
applied in order to obtain the cohomology groups of the horizontal complex
(\ref{lmp94}).

From the physical viewpoint, an extension of the class of
exterior forms to $\cQ^*_\infty$ enables us to concern
effective field theories whose Lagrangians involve derivatives of arbitrary
high order \cite{gom}

Here, we study $d$-, $d_V$- and $d_H$-cohomology of
the horizontal complex (\ref{lmp94}) on the infinite-order jet space
$J^\infty Y$  of an affine bundle $Y\to X$. Note that affine bundles provide a
standard framework in quantum field theory because almost all existent
quantization schemes deal with linear and affine quantities. Moreover, the De
Rham cohomology groups of an affine bundle
$Y\to X$ are equal to those of its base
$X$. Therefore, as we will see, the obstruction to the exactness of the
horizontal complex (\ref{lmp94}) lies only in exterior forms on $X$.
Since the BRST operator ${\bf s}$ eliminate these forms, the global descent
equations can be constructed though their right-hand sides are not equal to
zero. 

Moreover, 
we can restrict our consideration  to vector bundles $Y\to X$  without loss of
generality as follows.   Let $Y\to X$ be a smooth affine bundle modelled over a
smooth vector bundle
$\ol Y\to X$. 
A glance at the transformation law (\ref{55.21}) shows that $J^\infty Y\to X$
is an affine topological bundle modelled on the vector bundle $J^\infty
\ol Y\to X$. This affine bundle admits a global section
$J^\infty s$ which is the infinite-order jet prolongation of a global section
$s$ of
$Y\to X$. With $J^\infty s$, we have a homeomorphism
\be
\wh s_\infty: J^\infty Y \ni q \mapsto q -(J^\infty s)(\pi^\infty(q))\in
J^\infty\ol Y
\ee
of the topological spaces $J^\infty Y$ and $J^\infty\ol Y$,
together with an exterior algebra isomorphism
$\wh s_\infty^*:
\ol\cO^*_\infty\to\cO^*_\infty$.
Moreover, it is readily observed that the pull-back morphism $\wh
s_\infty^*$ commutes with the differentials $d$, $d_V$ and $d_H$.
Therefore, the differential algebras 
$\cQ^*_\infty$ and $\ol\cQ^*_\infty$ have the
same $d$-, $d_V$- and $d_H$-cohomology. 

Given a smooth vector bundle $Y\to X$, we will show the following. 
\begin{itemize}\begin{enumerate}
\item The De Rham cohomology groups of the differential algebra
$\cQ^*_\infty$ are isomorphic to those of the base $X$.
\item Its $d_V$-cohomology groups are trivial.
\item The $d_H$-cohomology groups of contact elements of the
algebra $\cQ^*_\infty$ are trivial.
\item The $d_H$-cohomology groups of its horizontal elements (i.e., cohomology
of the horizontal complex (\ref{lmp94})) coincide with the De Rham cohomology
groups of the base $X$.
\end{enumerate}\end{itemize}

Note that the results (i) and (ii) are also true for the differential algebra
$\cO^*_\infty$. The result (iii) takes place for an arbitrary smooth
fibre bundle $Y\to X$, and recovers that in Refs. \cite{ander0,ander}, obtained
by means of the Mayer-Vietoris sequence.

\section{Differential algebra $\cQ^*_\infty$}

Throughout the paper, smooth manifolds are
real, finite-dimensional, Hausdorff,
paracompact, and connected.

Given the surjective inverse system (\ref{5.10}), we have the
direct system (\ref{5.7'})  of ringed spaces
$(J^kY,\gO^*_k)$  whose structure sheaves
$\gO^*_k$ are sheaves of differential $\Bbb R$-algebras of exterior
forms on finite-order jet manifolds $J^kY$, and
$\pi^r_{r-1}{}^*$ are the pull-back morphisms.
Throughout, we follow the terminology of
Ref. \cite{hir} where by a sheaf is meant a sheaf bundle.  The
direct system (\ref{5.7'}) admits a direct limit
$\gQ^*_\infty$ which is 
a sheaf of differential exterior $\Bbb R$-algebras on
the infinite-order jet space
$J^\infty Y$. This direct limit exists in the category of 
sheaves of
$\Bbb R$-modules, and the direct limits of the familiar
operations on exterior forms provide $\gQ^*_\infty$ with a  differential
exterior algebra structure \cite{hart}.

Accordingly, we have the direct system (\ref{5.7})
of the structure algebras $\cO^*_k=\G(J^kY,\gO^*_k)$ of global sections of
sheaves $\gO^*_k$, i.e., $\cO^*_k$ are differential $\Bbb R$-algebras
of (global) exterior forms on finite-order jet manifolds
$J^kY$. As was mentioned above, the direct limit of (\ref{5.7}) is a
differential exterior $\Bbb R$-algebra
$(\cO^*_\infty, \pi^{\infty*}_k)$, isomorphic to the algebra of
all exterior forms on finite-order jet manifolds modulo the pull-back
identification.

The crucial
point is that the limit $\cO^*_\infty$ of the direct system (\ref{5.7}) of
structure algebras of sheaves $\gO^*_k$ fails to coincide with the structure
algebra $\cQ^*_\infty=\G(J^\infty Y,\gQ^*_\infty)$ of the limit
$\gQ^*_\infty$ of the direct system (\ref{5.7'}) of these sheaves. The sheaf
$\gQ^*_\infty$, by definition, is the sheaf of germs of local exterior forms
on finite-order jet manifolds. These local forms constitute a
presheaf $\gO^*_\infty$  from which the sheaf $\gQ^*_\infty$ is
constructed. It means that, given a section
$\f\in\G(\gQ^*_\infty)$ of $\gQ^*_\infty$ over an open subset $U\in J^\infty Y$
and any point
$q\in U$, there exists a neighbourhood $U_q$ of $q$ such that
$\f|_{U_q}$ is the pull-back of a local exterior form on some finite-order jet
manifold. However, $\gO^*_\infty$ does not coincide with the canonical
presheaf $\G(\gQ^*_\infty)$ of sections of the sheaf $\gQ^*_\infty$.

In particular, the $\Bbb R$-ring $\cQ^0_\infty$ is isomorphic to the
above mentioned ring of real locally pull-back  functions on
$J^\infty Y$. Indeed, any element of
$\cQ^0_\infty$ defines obviously such a function on $J^\infty Y$. Conversely,
the germs of any locally pull-back function $f$ on $J^\infty Y$ belong to
the sheaf $\gQ^*_\infty$, i.e., $f$ is a section of $\gQ^*_\infty$, and
different such functions $f$ and $f'$ are different sections of
$\gQ^*_\infty$.

There are obvious monomorphisms of algebras 
$\cO^*_\infty \to\cQ^*_\infty$ and presheaves $\gO^*_\infty
\to\G(\gQ^*_\infty)$.

For short, we
agree to call $\gQ^*_\infty$ (resp. $\cQ^*_\infty$) the sheaf (resp.
algebra) of locally pull-back exterior forms on $J^\infty Y$. The
exterior algebra operations and differentials $d$, $d_V$, $d_H$ are defined on
$\cQ^*_\infty$ just as on
$\cO^*_\infty$. At the same time, it should be emphasized again that elements
of the differential algebras 
$\cO^*_\infty$ and
$\cQ^*_\infty$ are not differential forms on $J^\infty Y$ in a rigorous
sense. Therefore, the standard theorems, e.g., the well-known De Rham theorem
(\cite{hir}, Theorem 2.12.3) can not be applied automatically to these
differential algebras.  

\section{De Rham cohomology}

Let $Y\to X$ be an arbitrary smooth fibre bundle.
We consider the complex of sheaves of $\cQ^0_\infty$-modules
 \beq
0\to \Bbb R\to
\gQ^0_\infty\op\longrightarrow^d\gQ^1_\infty\op\longrightarrow^d
\cdots
\label{lmp71}
 \eeq
on the infinite-order jet space $J^\infty Y$ and the corresponding 
infinite-order De Rham complex
 \beq
0\to \Bbb R\to
\cQ^0_\infty\op\longrightarrow^d\cQ^1_\infty\op\longrightarrow^d
\cdots
\label{5.13'}
\eeq
of locally pull-back exterior forms on $J^\infty Y$. 

Since locally pull-back
exterior forms fulfill the Poincar\'e lemma, the complex of sheaves
(\ref{lmp71}) is exact. Since the paracompact space $J^\infty Y$ admits a
partition of unity performed by elements of $\cQ^0_\infty$ \cite{tak2}, the
sheaves
$\gQ^r_\infty$ of
$\cQ^0_\infty$-modules are fine for all $r\geq 0$ \cite{book00,hir}. Then
they are acyclic, i.e., the cohomology groups $H^{>0}(J^\infty
Y,\gQ^r_\infty)$ of the paracompact space $J^\infty Y$ with coefficients in
the sheaf
$\gQ^r_\infty$ vanish \cite{hir}. Consequently, the exact sequence
(\ref{lmp71}) is a fine resolution of the constant sheaf $\Bbb R$ 
of germs of local constant real functions on
$J^\infty Y$. Then the well-known (generalized De Rham)
theorem on a resolution of a sheaf on a paracompact space (\cite{hir},
Theorem 2.12.1) can be called into play in order to find the cohomology
groups of the infinite-order De Rham complex complex (\ref{5.13}). 

In accordance with this
theorem, we have an isomorphism
\beq
H^*(\cQ^*_\infty)=H^*(J^\infty Y,\Bbb R) \label{lmp82}
\eeq 
of the De Rham cohomology groups $H^*(\cQ^*_\infty)$ of the
differential algebra $\cQ^*_\infty$ and the cohomology groups
$H^*(J^\infty Y,\Bbb R)$ of the infinite-order jet space $J^\infty Y$ with
coefficients in the constant sheaf $\Bbb R$. In the case of a vector bundle
$Y\to X$, we can say something more.

\begin{lem} \label{lmp72}  
If $Y\to X$ is a vector
bundle, there is an isomorphism 
\beq
H^*(J^\infty Y,\Bbb R)= H^*(X,\Bbb R) \label{lmp80}
\eeq 
of cohomology groups of
the infinite-order jet space $J^\infty Y$ with coefficients in
the constant sheaf $\Bbb R$ and those $H^*(X,\Bbb R)$ of the base $X$.
\end{lem}

\begin{proof}
The cohomology groups with coefficient in the constant sheaf $\Bbb R$ on
homotopic paracompact topological spaces are isomorphic \cite{span}. 
If $Y\to X$ is a vector bundle, its base $X$ is  
a strong deformation
retract of the infinite-order jet space $J^\infty Y$. To show this, let us
consider the map
\be
[0,1]\times J^\infty Y\ni (t; x^\la, y^i_\La) \to (x^\la, ty^i_\La)\in
J^\infty Y.
\ee
A glance at the transition functions (\ref{55.21}) shows that, given in the
coordinate form, this map is well-defined if $Y\to X$ is a vector bundle. 
It is a desired homotopy from $J^\infty Y$ to the base $X$ which is identified
with its image under the global zero section of the vector bundle $J^\infty
Y\to X$. 
\end{proof}

Combining the isomorphisms of cohomology groups (\ref{lmp82}), (\ref{lmp80}),
and the well-known isomorphism $H^*(X,\Bbb R)=H^*(X)$, we come to the
manifested isomorphism
\be
H^*(\cQ^*_\infty)=H^*(X) 
\ee
of the De Rham cohomology groups $H^*(\cQ^*_\infty)$ 
of the differential algebra
$\cQ^*_\infty$ of locally pull-back forms on the infinite-order jet
space $J^\infty Y$ of a vector bundle $Y\to X$ to the De Rham cohomology
groups
$H^*(X)$ of the base $X$.

It follows that any closed form $\f\in\cQ^*_\infty$ on $J^\infty Y$
is decomposed into the sum 
$\f=\varphi +d\xi$
where $\varphi\in \cO^*(X)$ is a closed form on $X$.

\section{Cohomology of $d_V$}

Due to the nilpotency rule (\ref{lmp50}), the vertical and horizontal
differentials $d_V$ and $d_H$ on the differential exterior algebra
$\cQ^*_\infty$ define the bicomplex 
\beq
\begin{array}{ccccrlcrlccrlccrlcc}
& & & & _{d_V} & \put(0,-10){\vector(0,1){20}} & & _{d_V} &
\put(0,-10){\vector(0,1){20}} & &  & _{d_V} &
\put(0,-10){\vector(0,1){20}} & & &  _{d_V} &
\put(0,-10){\vector(0,1){20}}& &\\ 
 &  & 0 & \to & &\cO^{k,0}_\infty &\ar^{d_H} & & \cQ^{k,1}_\infty &
\ar^{d_H} &\cdots  & & \cQ^{k,m}_\infty &\ar^{d_H} &\cdots & &
\cQ^{k,n}_\infty & & \\  
 & &  &  & & \vdots & & & \vdots  & & & 
&\vdots  & & & &
\vdots & &  \\ 
& & & & _{d_V} &\put(0,-10){\vector(0,1){20}} & & _{d_V} &
\put(0,-10){\vector(0,1){20}} & & &  _{d_V}
 & \put(0,-10){\vector(0,1){20}} & &  & _{d_V} & \put(0,-10){\vector(0,1){20}}
 & &\\
0 & \to & \Bbb R & \to & & \cQ^0_\infty &\ar^{d_H} & & \cQ^{0,1}_\infty &
\ar^{d_H} &\cdots  & &
\cQ^{0,m}_\infty & \ar^{d_H} & \cdots & &
\cQ^{0,n}_\infty & &\\
& & & & _{\pi^{\infty*}}& \put(0,-10){\vector(0,1){20}} & & _{\pi^{\infty*}} &
\put(0,-10){\vector(0,1){20}} & & &  _{\pi^{\infty*}}
 & \put(0,-10){\vector(0,1){20}} & &  & _{\pi^{\infty*}} &
\put(0,-10){\vector(0,1){20}} & & \\
0 & \to & \Bbb R & \to & & \cQ^0(X) &\ar^d & & \cQ^1(X) &
\ar^d &\cdots  & &
\cQ^m(X) & \ar^d & \cdots & &
\cQ^n(X) & \ar^d & 0\\
& & & & &\put(0,-10){\vector(0,1){20}} & & &
\put(0,-10){\vector(0,1){20}} & & & 
 & \put(0,-10){\vector(0,1){20}} & & &   &
\put(0,-10){\vector(0,1){20}} & & \\
& & & & &0 & &  & 0 & & & & 0 & & & & 0 & & 
\end{array}
\label{7}
\eeq
[5-8,10,11,13,14]. The rows and columns of these bicomplex are horizontal and
vertical complexes. Let us consider a vertical one
\beq
0\to \cQ^m(X) \ar^{\pi^{\infty*}} \cQ^{0,m}_\infty \ar^{d_V}\cdots \ar^{d_V} 
\cQ^{k,m}_\infty \ar^{d_V}\cdots, \qquad m\leq n. \label{lmp90}
\eeq

\begin{prop} \label{lmp53} 
If $Y\to X$ is a vector bundle, then the vertical complex (\ref{lmp90}) is
exact.
\end{prop}

\begin{proof}
Local exactness of a vertical complex on a coordinate chart
$((\pi^\infty)^{-1}(U_X); x^\la, y^i_\La)$, $0\leq|\La|,$ on $J^\infty Y$
follows from a version of the Poincar\'e lemma with
parameters (see, e.g., \cite{tul}). We have the the corresponding 
homotopy operator
\be
\si=\int^1_0 t^k [ \ol y\rfloor\f(x^\la, ty^i_\la)]dt, \qquad \f\in
\cQ^{k,m}_\infty,
\ee
where $\ol y=y^i_\La\dr_i^\La$. Since $Y\to X$ is a
vector bundle, it is readily observed that this homotopy operator is globally
defined on $J^\infty Y$, and so is the exterior form $\si$.
\end{proof}

It means that any
$d_V$-closed form $\f\in\cQ^*_\infty$ is the sum 
$\f=\varphi +d_V\xi$ of a $d_V$-exact form and an exterior form $\varphi$ 
on $X$.

Of course, $d_V$-cohomology of the bicomplex (\ref{7}) is not
trivial if the typical fibre of the fibre bundle $Y\to X$ is not contractible.

\section{Cohomology of $d_H$}

Turn now to the rows of the bicomplex (\ref{7}) (excluding the bottom one
which is obviously the De Rham complex on the base $X$). The
algebraic Poincar\'e lemma (see, e.g.,
\cite{tul,olver}) is obviously extended to elements of $\cQ^*_\infty$.

\begin{prop} If $Y\to X$ is a contractible fibre bundle $\Bbb R^{n+m}\to \Bbb
R^n$, the rows of the bicomplex (\ref{7}) are exact, i.e., they are always
locally exact.
\end{prop}
 
It follows that the the corresponding complexes of sheaves of
contact forms
\beq
 0\to \gQ^{k,0}_\infty \ar^{d_H}\gQ^{k,1}_\infty\ar^{d_H}\cdots  
\op\longrightarrow^{d_H} 
\gQ^{k,n}_\infty, \qquad k>0, \label{lmp91}
\eeq
and the above mentioned horizontal complex
\beq
0\to \Bbb R \to \gQ^0_\infty \ar^{d_H} \gQ^{0,1}_\infty
\ar^{d_H}\cdots  
\op\longrightarrow^{d_H} 
\gQ^{0,n}_\infty \label{lmp92'}
\eeq
are exact. Recall that, since $J^\infty Y$ is paracompact and admits a
partition of unity by elements of $\cQ^0_\infty$, all sheaves except the
constant sheaf $\Bbb R$ in the complexes (\ref{lmp91}), (\ref{lmp92'}) are
fine. However, the exact sequences
(\ref{lmp91}) and (\ref{lmp92'}) fail to be fine resolutions of the sheaves
$\gQ^{k,0}_\infty$ and $\Bbb R$, respectively, because of their last terms. At
the same time, following  directly the proof of the above mentioned
generalized De Rham theorem (\cite{hir}, Theorem 2.12.1) till these terms, one
can show the following.

\begin{prop} \label{lmp99'} If $Y\to X$ is an arbitrary smooth
fibre bundle, then the cohomology groups $H^r(k,d_H)$, $r<n$, of the complex 
of $\cQ^0_\infty$-modules 
\be
 0\to \cQ^{k,0}_\infty \ar^{d_H}\cQ^{k,1}_\infty\ar^{d_H}\cdots  
\op\longrightarrow^{d_H} 
\cQ^{k,n}_\infty
\ee
are
isomorphic to the cohomology groups $H^r(J^\infty Y,\gQ^{k,0})$ of $J^\infty
Y$ with coefficients in the sheaf $\gQ^{k,0}$ and, consequently, are trivial
because the sheaf $\gQ^{k,0}$ is fine.
\end{prop}

\begin{prop} \label{lmp99} 
If $Y\to X$ is an arbitrary smooth fibre bundle, 
the cohomology groups $H^r(d_H)$, $r<n$, of the horizontal complex
(\ref{lmp94}) are isomorphic to the
cohomology groups $H^r(J^\infty Y,\Bbb R)$ of $J^\infty Y$ with
coefficients in the constant sheaf $\Bbb R$.
\end{prop}

Note that one can also study the exact sequence of presheaves
\be
0\to \Bbb R_\infty\to
\gO^0_\infty\op\longrightarrow^d\gO^1_\infty\op\longrightarrow^d
\cdots,
\ee
but comes again to the results of Propositions \ref{lmp99'}, \ref{lmp99}.
Because $J^\infty Y$ is paracompact, the cohomology groups $H^*(J^\infty Y,
\gQ^*_\infty)$ of
$J^\infty Y$ with coefficients in the sheaf $\gQ^*_\infty$ and those
$H^*(J^\infty Y, \gO^*_\infty)$ with coefficients in the presheaf
$\gO^*_\infty$ are isomorphic. It follows that the cohomology group
$H^0(J^\infty Y, \gO^*_\infty)$ of the
presheaf $\cO^*_\infty$ is isomorphic to the $\Bbb R$-module
$\cQ^*_\infty=H^0(J^\infty Y, \gQ^*_\infty)$, but not
$\cO^*_\infty$.

If $Y\to X$ is a vector bundle, Lemma \ref{lmp72} and Proposition \ref{lmp99}
lead to the manifested isomorphism
\be
H^r(d_H)=H^r(X,\Bbb R)= H^r(X), \qquad r<n,
\ee
of the $d_H$-cohomology groups $H^{<n}(d_H)$ of the horizontal 
complex (\ref{lmp94}) to the De Rham cohomology groups $H^{<n}(X)$ of the base
$X$. 
Then, combining Propositions \ref{lmp99'} and \ref{lmp99}, we conclude that 
any
$d_H$-closed form
$\f\in
\cQ^*_\infty$ on the infinite-order jet manifold $J^\infty Y$ is decomposed
into the sum 
\beq
\f=\varphi +d_H\xi \label{lmp101}
\eeq
 where $\varphi\in \cO^*(X)$ is a
closed form on $X$. 

Turn to outcomes of this result to BRST
theory. Since
${\bf s}\varphi=0$, the decomposition (\ref{lmp101}) prevents one from the
topological obstruction to definition of global descent equations in BRST
theory on vector (and affine) bundles.


\begin{thebibliography}{ederf}


\bibitem{henn91} M.Henneaux, Space-time locality of the BRST formalism, {\sl
Commun. Math. Phys.}, {\bf 140}, 1-13 (1991).

\bibitem {barn} G.Barnish, F.Brandt and M.Henneaux, Local BRST cohomology in
the antifield formalism. 1. General theorems, {\sl Commun. Math. Phys.}, {\bf
174}, 57-91 (1995).

\bibitem{brandt} F.Brandt, Local BRST cohomology and covariance, {\sl Commun.
Math. Phys.}, {\bf 190}, 459-489 (1997).

\bibitem{book00} L.Mangiarotti and G.Sardanashvily, {\it Connections in
Classical and Quantum Field Theory}, World Scientific, Singapore, 2000.

\bibitem{tak2} F.Takens, A global version of the inverse problem of
the calculus of variations, {\it J. Diff. Geom.}, {\bf 14}, 543-562 (1979).

\bibitem{bau} M.Bauderon, Differential geometry and Lagrangian formalism in
the calculus of variations, in {\it Differential Geometry, Calculus of
Variations, and their Applications}, Lecture Notes in Pure and Applied
Mathematics, {\bf 100}, Marcel Dekker, Inc., N.Y., 1985, pp. 67-82.

\bibitem{tak1} F.Takens, Symmetries, conservation laws and variational
principles, in {\it Geometry and Topology}, Lect. Notes in Mathematics, {\bf
597}, Springer-Verlag, Berlin, 1977, pp. 581-604.

\bibitem{book} G.Giachetta, L.Mangiarotti and G.Sardanashvily, {\it New
Lagrangian and Hamiltonian Methods in Field Theory}, World Scientific,
Singapore, 1997.

\bibitem{abb} M.Abbati and A.Mani\`a, On differential structure for
projective limits of manifolds, {\it J. Geom. Phys.}, {\bf 29}, 35-63 (1999).

\bibitem{ander0} I.Anderson, Introduction to the variational bicomplex, {\it
Contemp. Math.}, {\bf 132}, 51-74 (1992).

\bibitem{ander} I.Anderson, {\it The Variational Bicomplex}, Academic Press,
Boston, 1994.

\bibitem{jpa00} G.Giachetta, L.Mangiarotti and G.Sardanashvily, On the
obstruction to the exactness of the variational bicomplex, arXiv:
math-ph/0004024.

\bibitem{tul} W.Tulczyjew, The Euler--Lagrange resolution, 
in {\it Differential
Geometric Methods in Mathematical Physics}, Lect. Notes in Mathematics,
{\bf 836} (Springer-Verlag, Berlin, 1980), pp. 22-48.

\bibitem{olver} P.Olver, {\it Applications of Lie Groups to Differential
Equations}, Graduated Texts in Mathematics, {\bf 107}, Springer-Verlag,
Berlin, 1986.

\bibitem{hir} F.Hirzebruch, {\it Topological Methods in Algebraic Geometry}
(Springer-Verlag, Berlin, 1966).

\bibitem{gom} J.Gomis and S.Weinberg, Are nonrenormalizable gauge theories
renormolizable?, {\it Nucl. Phys.}, {\bf B469}, 473-487 (1996).

\bibitem{hart} R.Hartshorne, {\it Algebraic Geometry}, Graduate Texts in
Mathematics, {\bf 52}, Springer-Verlag, Berlin, 1977.

\bibitem{span} E.Spanier, {\it Algebraic Topology}, McGraw-Hill Book Company,
N.Y., 1966.

\end{thebibliography}
\end{document}